\documentclass[prl,twocolumn]{revtex4}
\usepackage{amssymb}
\usepackage{graphicx}
\usepackage{dcolumn}
\usepackage{bm}

\begin{document}

\title{Superadditivity of distillable entanglement From Quantum
Teleportation }
\author{Somshubhro Bandyopadhyay\footnote{Present address: Institute for Quantum Information Science, University of Calgary, Calgary AB, Canada}}
\email{som@qis.ucalgary.ca}
\affiliation{Centre for Quantum Information and Quantum Control and Department of Chemistry, University of Toronto, Toronto, ON M5S 3H6, Canada.}
\author{Vwani Roychowdhury}
\email{vwani@ee.ucla.edu}
\affiliation{Electrical Engineering Department, UCLA, Los Angeles, CA 90095}

\begin{abstract}
{\small We show that the phenomenon of superadditivity of distillable
entanglement observed in multipartite quantum systems results from the
consideration of states created during the execution of the standard
end-to-end quantum teleportation protocol (and a few additional local
operations and classical communication (LOCC) steps) on a linear chain of
singlets. Some of these intermediate states are tensor products of
bound-entangled (BE) states, and hence, by construction possess distillable
entanglement, which can be unlocked by simply completing the rest of the
LOCC operations required by the underlying teleportation protocol. We use
this systematic approach to construct both new and known examples of
superactivation of bound entanglement, and first examples of activation of
BE states using other BE states. A surprising outcome is the construction of
noiseless quantum relay channels with no distillable entanglement between
any two parties, except for that between the two end nodes.}
\end{abstract}

\maketitle

%\date{ }

Quantum entanglement is fundamental to quantum
mechanics, and has been established as the critical resource for quantum
information processing (QIP) and quantum computing. In spite of much recent progress, spurred by the growth of the new field
of QIP, a comprehensive functional and conceptual understanding of
entanglement continues to remain elusive. In this letter, we explore one
such intriguing issue involving the emergence of distillable entanglement in
a multi-party quantum system, when none of the constituent subsystems
possesses any distillable entanglement.

More precisely, an entangled state is said to be distillable if one can
obtain some pure entanglement in an asymptotic sense by LOCC \cite{distill}.
This class of entangled states can be used to set up quantum channels and
provide the basic infrastructure for communication and computation. Even
though most entangled states are distillable, there exist entangled states
that do not allow distillation of maximally entangled states \ by LOCC. Such
states are known as bound entangled (BE) states \cite{boundent}. A
multiparty entangled state is said to be bound entangled if there is no
distillable entanglement between any subset of the parties \emph{as long as
all the parties remain separated from each other}. Recent results have shown
that bound entangled states in multipartite systems come in two varieties:
the activable (or unlockable) and non-activable states. For activable BE
(ABE) states \cite{Dur,smolin}, if some of the parties group together and
perform collective LOCC, then they can distill entanglement between a subset
of spatially separated parties. However, not all multiparty bound entangled
states are activable, and there exist states where there is no distillable
entanglement across any bipartite partition. An example of such a state is
the bound entangled state constructed from a multiparty unextendible product
basis \cite{upb}.

An intriguing example of the richness of entanglement manipulation in a
multiparty scenario is the so-called superactivation process, where two
activable bound entangled states tensored together produces distillable
entanglement between two parties belonging to the two \textit{different}
states \cite{sst}. It demonstrates superadditivity of distillable
entanglement as the individual BE states are not distillable. At this point
it is useful to make a \emph{distinction} between \emph{superactivation} and 
\emph{activation}. In activation, one is provided with a multiparty BE state
such that if some of the parties are provided with {\em additional states} to
share, then a subset of the spatially separated parties can distill entanglement
among them. In the literature, the shared auxiliary states have been
singlets \cite{Dur,smolin}, e,g., in an ABE state, bringing a subset of the
parties together is equivalent to providing them with singlets. Moreover, 
in certain multipartite activation
processes, the provided singlet is entangled across a different
cut than the entanglement obtained in activation, and as such, this information should be always explicitly
stated.  An important point 
to note is that all 
the activation processes {\em demonstrated show far} correspond to the {\em subadditive property} of distillable entanglement: In order to distill one maximally entangled state
from the given state, one, on the average, has to spend more than one maximally entangled states. One of the results we show is that one can also unlock certain ABE
states by providing another BE state, instead of singlets; thus, it is
possible to \emph{demonstrate superadditivity of distillable entanglement in
the context of the activation process} as well.

What is the physics behind the superadditivity of distillable entanglement
in a multiparty situation? To answer this, it is useful to consider the
particular four-party ABE state that lies at the center of the
superactivation process \cite{smolin,sst}: 
\begin{equation}
\rho _{s}^{ABCD}=\frac{1}{4}\sum_{i=1}^{4}\left( \left\vert \Phi
_{i}\right\rangle \left\langle \Phi _{i}\right\vert \right) ^{AB}\otimes
\left( \left\vert \Phi _{i}\right\rangle \left\langle \Phi _{i}\right\vert
\right) ^{CD},  \label{eq:ube state}
\end{equation}%
where for convenience sake, we have adopted the following notation for the
four Bell states: $\Phi =\left\{ \Psi ^{+},\Psi ^{-},\Phi ^{+},\Phi
^{-}\right\} $ with elements $\Phi _{i}$ being the maximally entangled
states for two qubits (Bell states) and are given by: $\displaystyle%
\left\vert \Psi ^{\pm }\right\rangle =\frac{1}{\sqrt{2}}\left( \left\vert
01\right\rangle \pm \left\vert 10\right\rangle \right) ,\left\vert \Phi
^{\pm }\right\rangle =\frac{1}{\sqrt{2}}\left( \left\vert 00\right\rangle
\pm \left\vert 11\right\rangle \right) $. The four-party ABE state, $\rho
_{s}$, has several interesting properties (for details see Ref \cite{smolin}%
): it is symmetric under interchange of parties, i.e., $\rho
_{s}^{ABCD}=\rho _{s}^{ACBD}=\rho _{s}^{ADCB}$ etc., and any two parties
coming together can distill a singlet between the other two which implies
the state must be entangled. On the other hand since the state is separable
across every 2:2 bipartite cut every party is separated from every other by
a separable partition and hence no entanglement can be distilled when all
four parties remain spatially separated. Hence the state is bound entangled.

We first show how to optimally prepare the state $\rho_s^{ABCD}$ and then
show how it can efficiently replace a pair of singlets in a relay quantum
channel; \emph{this ability of the ABE state to replace singlets} in a
chain, is the key to the phenomenon of superadditivity of distillable
entanglement.
\vspace*{-6pt}
\begin{center}
\textbf{Optimal Preparation of the ABE state $\rho_s^{ABCD}$ in Eq.~(\ref%
{eq:ube state})}
\end{center}
\vspace*{-6pt}
Recently it was shown that the Smolin state belong to the family of
activable bound entangled states for even number (greater than or equal to
four) of qubits \cite{BCRS04} and the exact cost of preparation of such
states have been shown to be \textit{N} ebits for a \textit{2N} qubit ABE
state \cite{BR04}. In particular it was shown that \emph{two ebits
are both necessary and sufficient} to prepare the four qubit Smolin state  \cite{BR04}. We now briefly review the sufficiency part of the proof following  \cite{BR04}. The sufficiency part of the proof involves a protocol 
that utilizes a pair of singlets (hence, requiring two ebits) and LOCC. In fact throughout this paper we will make use of this local protocol. Let the pairs, (A, B), and (C, D), share a singlet each. A and C can classically communicate among
themselves to prepare a state $\left\vert \Phi_{i}\right\rangle^{AA} \otimes
\left\vert \Phi_{i}\right\rangle^{BB} $ randomly with equal probability. A
and C then can each teleport one qubit of the correlated Bell states
(keeping one qubit from each state to themselves) to B and D using the
shared singlets. This creates the ABE state $\rho_s^{ABCD}$ (see Fig. 1).

\begin{figure}[tbp]
\includegraphics[scale=0.5]{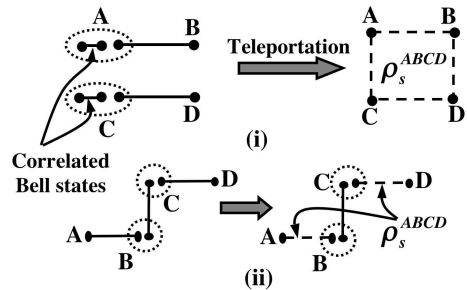}
\caption{(i) Optimal preparation of the ABE state $\protect\rho_s^{ABCD}$.
Singlets and other Bell states are shown by solid lines connecting filled
circles. Multipartite states are represented by dotted lines connecting the
underlying parties. (ii) Teleportation in a chain of singlets and in a system
consisting of one ABE state and a singlet.}
\label{fig:1}
\end{figure}

The construction of the state makes it clear why the ABE state $\rho
_{s}^{ABCD}$ can replace a pair of singlets in a chain. As illustrated in
Fig. 1(ii), consider the case where parties A, B, C, and D are linked by
three singlets. It is obvious that one can generate a singlet state between
the end nodes A and D just by following the standard teleportation protocol
in the intermediate nodes via the process of entanglement swapping. Now
suppose the four parties conspire and replace the pair of singlets AB and CD
with the ABE state $\rho _{s}^{ABCD}$. The state of the quantum channel is
now $\rho _{s}^{ABCD}\otimes (\left\vert \Psi ^{-}\right\rangle \left\langle
\Psi ^{-}\right\vert )^{BC}$. Next, \emph{if this modified state is given to
four new parties}, then they \emph{can treat the chain as if it were still
singlets}, and distill one ebit of entanglement by executing the \emph{%
standard teleportation} protocol. To see this, consider the situation where
A receives a one-qubit state $\rho $. The teleportation from A to B happens
via an unknown Bell state, and hence the teleported state in B is $\sigma
_{i}\rho \sigma _{i}^{\dag }$, for some unknown Pauli operator $\sigma_i$.
The state remains the same in C after teleportation from B to C as the
channel is a singlet. However in the last step of teleportation from C to D,
it takes place using the same but unknown Bell state as that between A and
B. Hence, the final state becomes $\sigma _{i}^{2}\rho \sigma _{i}^{2\dag }$%
; as $\sigma _{i}^{2}=I,$ \emph{the final state at node D is again }$\rho$.%
\newline

\noindent  \textbf{Remark 1:} \emph{These arguments also show why $%
\rho_s^{ABCD}$ is activable}: providing a singlet between B and C is the
same as bringing the pair together; hence, by bringing B and C together we
can distill an ebit between A and D. In other words, there is one ebit of
distillable entanglement across any 1:3 cut in $\rho_s^{ABCD}$.\newline
\textbf{Remark 2:} While we have presented the case of a chain with three
singlets, the arguments easily generalize to the case of chains of singlets
of any length: \emph{one can distill one ebit of entanglement} between the
two end parties, by performing \emph{the standard teleportation protocol},
even if any number of \emph{pairs of singlets are first converted to the ABE
state} $\rho_s^{ABCD}$.

The properties of the state $\rho _{s}^{ABCD}$ are, however, indeed
different from that of a pair of singlets. To see this, consider the two
qubit channels formed by a pair of singlets, and by the four-party state $
\rho _{s}^{ABCD}$. Given any 2-qubit state $\rho $ at one end of the
channel, the teleported state on the other side is $\frac{1}{4}%
\sum_{i=1}^{4}(\sigma _{i}\otimes \sigma _{i})\rho (\sigma _{i}\otimes
\sigma _{i})$, which will be same as the input state, $\rho $, only for
special states. For example 2-qubit states that are invariant under $\sigma
_{i}\otimes \sigma _{i}$ operations, can be transmitted exactly over the ABE
channel. For instance, Werner states are indeed invariant under such an
operation and hence, they can be teleported unaltered via the 2-qubit
channel $\rho _{s}^{ABCD}$.

\vspace*{-6pt}
\begin{center}
\textbf{Superactivation from Teleportation }
\end{center}
\vspace*{-6pt}
Consider the chain of seven singlets, as shown in Fig. 2. As the figure
illustrates, two pairs of singlets can be replaced with their corresponding
four-party ABE states, $\rho _{s}$. \emph{It follows from Remark 2} that the
resulting state has one ebit of entanglement between A and E, which can be
recovered by the standard teleportation protocol. However, as shown in the
figure, if one uses the remaining three singlets to bring the respective
parties together, then the chain reduces to the tensor product state: $\rho
_{s}^{ABCD}\otimes \rho _{s}^{EBCD}$. This is the case of \emph{%
superactivation} introduced in \cite{sst}: the tensor product of two ABE
states leads to distillable entanglement between two parties belonging to
the two different states.

\begin{figure}[tbp]
\includegraphics[scale=0.5]{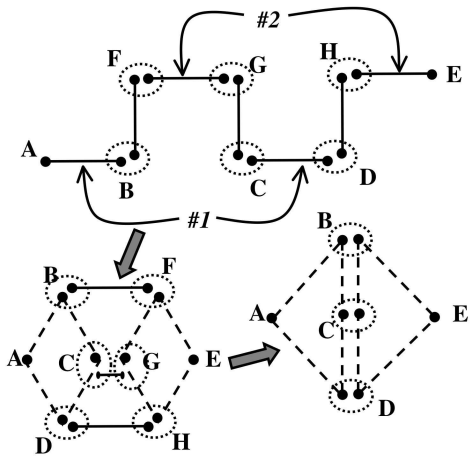}
\caption{Superactivation involving two four-party activable bound-entangled
states (ABEs), as constructed from a singlet chain via LOCC and the standard
teleportation protocol.}
\label{fig:2}
\end{figure}

As shown in Fig. 3, one could replace a third pair (e.g., the singlets BF
and DH) in the original 7-singlet chain by the corresponding ABE state. 
\emph{It follows again from Remark 2} that the resulting state still has one
ebit of entanglement between A and E, which can be distilled by the standard
teleportation protocol. Next, if one brings C and G together by using the
singlet CG, then one gets a \emph{new superactivation scenario}: The tensor
product state, $\rho _{s}^{ABCD}\otimes \rho ^{BDFH}\otimes \rho _{s}^{CEFH}$%
, of three ABE states leads to one ebit of distillable entanglement between
the nodes A and E. Clearly, one can now create infinitely many such
superactivation configurations.

\begin{figure}[tbp]
\includegraphics[scale=0.5]{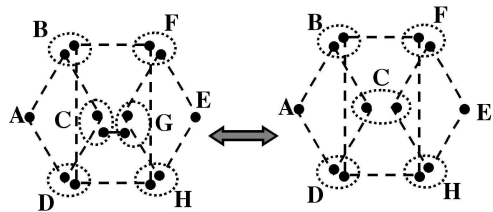}
\caption{Superactivation involving three ABE states.}
\label{fig:3}
\end{figure}

\noindent \textbf{Remark 3:} In the original superactivation configuration,
as illustrated in Fig. 2, the state is the tensor product of two \emph{%
disjoint} ABE states connected via three singlets: $\rho _{s}^{ABCD}\otimes
\rho _{s}^{EFGH}\otimes (\left\vert \Psi ^{-}\right\rangle \left\langle \Psi
^{-}\right\vert )^{BF}\otimes (\left\vert \Psi ^{-}\right\rangle
\left\langle \Psi ^{-}\right\vert )^{CG}\otimes (\left\vert \Psi
^{-}\right\rangle \left\langle \Psi ^{-}\right\vert )^{DH}$. Can the two ABE
states be activated with less than three singlets, or equivalently, by
sharing less than three parties between the two ABE states? Going back to
the original 7-singlet chain, one can easily show that (i) the original
chain will break up into at least two pieces, if one or more of the three
singlets are removed, and (ii) each connected chain will have only one of
the two correlated Bell states from at least one of the ABE states. For
example, if the singlet BF is removed in Fig. 2, then the chain breaks into
two, and the two pairs, AB and CD, from the state $\rho _{s}^{ABCD}$, are in
different connected chains. Because of these unmatched correlated Bell
states, each chain acts as a depolarizing channel; distillable entanglement
is obtained only if both the pairs of the ABE state $\rho _{s}$ are in the
same connected chain. Hence, two ABE states, $\rho _{s}$, must share three
parties in order to have distillable entanglement between A and E. In other
words, \emph{any state comprising two such states $\rho _{s}$ is a ABE state
if only two or less parties are shared between the two states}.

\vspace*{-6pt}
\begin{center}
\textbf{Superadditivity of Distillable Entanglement in the Activation
Scenario }
\end{center}
\vspace*{-6pt}
Consider the superactivation configuration involving three ABE states: $\rho
_{s}^{ABCD}\otimes \rho _{s}^{BDFH}\otimes \rho _{s}^{CEFH}$ (see Fig. 3).
As already mentioned, in order to distill one ebit of entanglement between A
and E, the rest of the nodes can follow the standard teleportation protocol,
where the nodes can perform their respective Bell measurements and announce
the results to one of their neighbors \emph{in any order} (follows from
Remark 2 and the commuting properties of Bell measurements). Consider the
case where the node C performs its Bell measurement (BM) and announces its
results. After this, the state becomes the tensor product of one six-party
state and one four-party state: $\rho _{x}^{ABDEFH}\otimes \rho _{s}^{BDFH}$%
. Since one can still distill one ebit of entanglement between A and E by
having the rest of the nodes complete their BM's and classical
communications (CC's), the new configuration, $\rho _{x}^{ABDEFH}\otimes
\rho _{s}^{BDFH}$, is distillable.

We have already shown in Remark 3 that the configuration $\rho
_{s}^{ABCD}\otimes \rho _{s}^{CEFH}$ (i.e., two ABE states $\rho _{s}$,
where only one party is shared between the two ) is ABE, and the state $\rho
_{x}^{ABDEFH}$ is obtained from the two states via LOCC; hence, \emph{the
state $\rho _{x}^{ABDEFH}$ is also a ABE state}.

Thus, the distillable state $\rho_x^{ABDEFH}\otimes \rho_s^{BDFH}$ can be
viewed as follows: One is given a six-party BE state $\rho_x^{ABDEFH}$,
which is unlocked by using another BE state, $\rho_s^{BDFH}$, as an
auxiliary resource. Thus \emph{the BE state $\rho_x^{ABDEFH}$ is activated},
not by bringing parties together or by giving free entanglement, but \emph{%
by another BE state}. This constitutes a case of superadditivity of
distillable entanglement in the activation scenario. Note that we obtained
the activation case from the superactivation case by performing one of the
BM's necessary in the distillation process. This reiterates our thesis that
the superadditivity phenomenon in multipartite systems is just a
manifestation of the distillation process via teleportation.

\begin{figure}[tbp]
\includegraphics[scale=0.5]{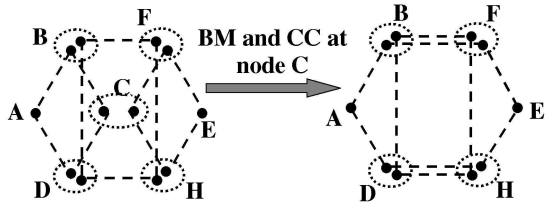}
\caption{Superadditivity of distillable entanglement in the activation
framework.}
\label{fig:4}
\end{figure}

\vspace*{-6pt}
\begin{center}
\textbf{Discussions }
\end{center}
\vspace*{-6pt}
We have shown that all the known and several new instances of the phenomena of superadditivity of distillable
entanglement in multipartite quantum systems can be systematically derived
by performing {\em the standard quantum teleportation or entanglement-swapping 
protocol} in a chain of singlets. 
One might ask if the same mechanism would hold for {\em any} case of superactivation
that might be discovered in the future.   It is a well-known truism in QIP that given enough pairwise ebits,  {\em any} multipartite state can be prepared via LOCC. For example, if one party, say $A$, shares one ebit with every other party then this configuration guarantees preparation of any multipartite state; such a configuration is referred to as a {\em star} network, with $A$ as the central hub. This is because, $A$ can prepare the multipartite state, and teleport corresponding qubits to each of the other parties. Equivalently, starting from a chain of sufficient number of ebits, the above star configuration can always be generated by LOCC. Next, one can always replace the ebits by sufficient numbers of singlets, leading to the conclusion that a nonuniform chain of singlets (where we allow neighboring nodes to share more than one singlet) can generate any superactivation configuration via LOCC. Thus, \iffalse {\em in an almost axiomatic sense},\fi a nonuniform {\em chain of singlets} is {\em a universal configuration,} and can lead to any superactivation mechanism.  The appealing aspects of the results
in this paper are that (i) the chain of singlets is uniform and has exactly one singlet between neighboring nodes, and (ii) after the initial LOCC operations to set up the ABE states, the superactivation configurations can be constructed by exclusively executing 
the standard teleportation protocol. This provides the {\em first physical explanation} and a constructive procedure for a {\em phenomenon that was originally presented and perceived by the community as a puzzling and surprising} aspect of quantum entanglement.

We end the communication by considering a rather simple question: Suppose we want a
quantum relay channel connecting nodes A and E, and going through nodes B,
C, and D. 
\iffalse
The added restriction is that once the quantum state involving the
parties is prepared, the nodes can communicate classically only with its
neighbors:\newline
$A\leftarrow \hspace*{-2mm}\rightarrow B\leftarrow \hspace*{-2mm}\rightarrow
C\leftarrow \hspace*{-2mm}\rightarrow D\leftarrow \hspace*{-2mm}\rightarrow
E $.\fi 
Is it necessary that the pairs of
neighboring nodes, (A,B), (B,C), (C,D), and (D,E) must have distillable
entanglement? Quite unexpectedly, we find that the answer is in the negative.
Consider a channel comprising a linear chain of four singlets; clearly, the
neighboring nodes in the chain are maximally entangled. Now, however, we use
the singlets, AB and CD, to prepare (via LOCC) the ABE state $\rho
_{s}^{ABCD}$, and the singlets, BC and DE, to prepare the ABE state $\rho
_{s}^{BCDE}$. This leads to the superactivation configuration, $\rho
_{s}^{ABCD}\otimes \rho _{s}^{BCDE}$, from which 
one can distill one ebit of
entanglement between A and E. Thus, the
state $\rho _{s}^{ABCD}\otimes \rho _{s}^{BCDE}$ comprises a noise-less
quantum relay channel, but with a \emph{surprising twist}: if one considers
any pair of nodes in the channel (including, the pairs formed by the
neighboring nodes), except the end pair (A, E), then there is no distillable
entanglement between the parties in the pair!

\noindent {\large \emph{Acknowledgements}}\newline
This work was sponsored in part by the U. S. Army
Research Office/DARPA under contract/grant number DAAD 19-00-1-0172, and in 
part by the NSF grant CCF:0432296.

\end{document}